# Plastic-crystalline solid-state electrolytes: Ionic conductivity and orientational dynamics in nitrile mixtures

D. Reuter, P. Lunkenheimer[a)], and A. Loidl

**AFFILIATION**

Experimental Physics V, Center for Electronic Correlations and Magnetism, University of Augsburg, 86135 Augsburg, Germany

[a)] **Electronic mail:** peter.lunkenheimer@physik.uni-augsburg.de

**ABSTRACT**

Many plastic crystals, molecular solids with long-range, center-of-mass crystalline order but dynamic disorder of the molecular orientations, are known to exhibit exceptionally high ionic conductivity. This makes them promising candidates for applications as solid-state electrolytes, e.g., in batteries. Interestingly, it was found that the mixing of two different plastic-crystalline materials can considerably enhance the ionic dc conductivity, an important benchmark quantity for electrochemical applications. An example is the admixture of different nitriles to succinonitrile, the latter being one of the most prominent plastic-crystalline ionic conductors. However, until now only few such mixtures were studied. In the present work, we investigate succinonitrile mixed with malononitrile, adiponitrile, and pimelonitrile, to which 1 mol% of Li ions were added. Using differential scanning calorimetry and dielectric spectroscopy, we examine the phase behavior and the dipolar and ionic dynamics of these systems. We especially address the mixing-induced enhancement of the ionic conductivity and the coupling of the translational ionic mobility to the molecular reorientational dynamics, probably arising via a "revolving-door" mechanism.

## I. INTRODUCTION

Electrolytes, materials where the electrical current is not carried by electrons but by ions, are essential for a number of electrochemical applications, including energy-storage and -conversion devices as batteries, fuel cells, or supercapacitors. Finding better electrolytes is the key for ensuring a sustainable energy supply of tomorrow as the non-continuous nature of solar and wind energy and the future success of electromobility require significant advances in energy-storage technologies. Therefore, the development and investigation of new electrolytes is one of the most important goals of current material science.

There are various strategies to achieve further progress in electrolyte technology. Maybe the most promising, but also most demanding approach is the use of solid materials with high ionic conductivity like certain crystalline compounds[1,2,3,4,5] or amorphous solids, e.g., polymers.[6,7,8] This avoids the shortcomings of many of the currently employed, mostly liquid electrolytes, as flammability, limited electrochemical stability, and leakage.[9] A recent development in this field (apart of pioneering works like Ref. 10) are ionically conducting plastic crystals (PCs), which were shown to reach conductivity values in the technically relevant range of $>10^{-4}\ \Omega^{-1}\ cm^{-1}$ at room temperature.[1,2,5,11,12,13,14,15,16,17,18] In PCs, the centers of mass of the molecules are located on a crystalline lattice, but they are dynamically disordered with respect to their orientational degrees of freedom, the latter often showing glassy freezing at low temperatures.[19] Obviously, there is only small hindrance for reorientational processes in these materials, indicating relatively weak mutual interactions between the PC molecules. This usually causes rather high plasticity (explaining the term "plastic crystals"), which helps avoiding junction problems to the metallic electrodes that may occur for other solid electrolytes.[18] To ensure the presence of technically relevant cations like $Li^+$, needed, e.g., for lithium batteries, usually some mol% of salts like $LiPF_6$ are added to the pure PC systems.

There are essentially two classes of plastic-crystalline electrolytes: ionic PCs, i.e. salts where the cation or the anion (or both) have orientational degrees of freedom,[1,5,11,12,15,17,20,21] and non-ionic PCs, formed by neutral molecules.[2,13,14,16,22] Interestingly, for two prominent examples of both material classes, it was found that admixing a related molecular species of larger size can considerably enhance the ionic conductivity and the stability range of the plastic phase: Within the *closo*-carborane related group of ionic PCs,[15,17] very recently a mixture of $Li^+(CB_9H_{10})^-$ with $Li^+(CB_{11}H_{12})^-$ was reported to exhibit high room-temperature conductivity and an extended PC phase compared to the pure compound.[5] For succinonitrile [SN; $C_2H_4(CN)_2$],[2,13] mixtures with glutaronitrile [GN; $C_3H_6(CN)_2$] led to a marked enhancement of the ionic conductivity in its PC phase,[16,23,24] reaching up to three decades for samples with admixed Li salt.[16]

Succinonitrile is an important representative of non-ionic PC electrolytes. Its high ionic conductivity when introducing ions by adding selected salts was revealed by Long et al.[13] and Alarco et al.[2] Admixing GN leads to a considerable extension of the plastic-crystalline state towards low temperatures.[25,26] This enables the observation of glassy freezing of the orientational disorder below about 150 K, leading to a so-called glassy-crystal state (sometimes also termed "orientational glass"). In an earlier work,[16] we have reported dielectric-spectroscopy measurements of SN-GN mixtures with different mixing ratios of up to 80 mol% GN. These experiments revealed successively stronger coupling of molecular reorientation and ionic translation with increasing GN content. This finding indicates that the mentioned conductivity enhancement of the mixtures can be explained by a "revolving door" or "paddle wheel" mechanism[10,11,27] which becomes optimized when part of the smaller SN molecules is replaced by the larger GN molecules.[16] Within this scenario, the molecule reorientations are assumed to provide transient free volume within the lattice, thus enhancing the ionic mobility. Quite generally, it is often assumed that the typical reorientational motions of the molecules or ions in PCs facilitate translational ionic hopping, thus explaining the high ionic conductivity of many members of this material class.[10,11,12,18,27,28,29,30] Recently, reorientational motions were even suggested to also play a role for the ionic mobility in certain liquid electrolytes.[31,32] However, this issue still is far from being finally clarified and, as an alternative, various types of defects occurring in the plastic-crystalline phase were also considered to rationalize the high ionic mobility.[1,2,14,20,33,34] Obviously, more experimental data on the coupling of ionic conductivity and the reorientational motions in mixed PC systems are needed to achieve a better understanding of the mechanisms governing the ionic mobility in PCs.

While in Ref. 16, we have only investigated the admixture of a single compound (GN) to SN, using other nitriles of different size seems a reasonable way to further explore the role of the revolving-door mechanism for the ionic conductivity. For example, adiponitrile [$C_4H_8(CN)_2$; AN] has one additional $CH_2$ group compared to GN. Therefore, following the arguments in Ref. 16, for an SN-AN mixture one may expect an even more effective revolving-door mechanism and stronger coupling of translational ionic and reorientational molecular motions. An alternative approach is the mixture of malononitrile [$CH_2(CN)_2$; MN] with SN. Here MN, having smaller molecules, may take the role of SN in the SN-GN mixtures and the bigger SN itself could act as optimizing material, widening the lattice and making revolving doors more effective for ionic charge transport. The miscibility and stabilization of the PC phase against the transition into the orientationally ordered crystal was already confirmed for the SN-MN system by unpublished work of List and Angell in 1985 (see remark in Ref. 18).

In the present work, we provide a detailed study of the ionic conductivity and reorientational dynamics of binary mixtures of SN with MN, AN, and pimelonitrile [$C_5H_{10}(CN)_2$; PN] using dielectric spectroscopy. In particular, we have investigated $SN_{0.85}X_{0.15}$ mixtures (with $X$ = MN, AN, or PN) where 1% $LiPF_6$ was added to enable a direct comparison to the results of our previous report[16] where the same salt admixture was used. This leads to the overall composition $(SN_{0.85}X_{0.15})_{0.99}(LiPF_6)_{0.01}$. It should be noted that the present study does not aim at achieving exceptionally high values of ionic conductivity (here different salts and higher concentrations may be more favorable[2]) but, instead, intends to contribute to a better understanding of the fundamental aspects of ion conduction in such mixed PC systems.

## II. EXPERIMENTAL METHODS

The investigated nitriles MN, SN, GN, AN (Arcos Organics) and PN (Sigma-Aldrich) were obtained with purities > 99%. $LiPF_6$ (purity 99.99%) was purchased from Sigma-Aldrich. Suitable amounts of the nitriles and $LiPF_6$ powder were mixed under slight heating to achieve complete dissolution. For differential scanning calorimetry (DSC) measurements, a DSC 8500 from Perkin Elmer was used applying a constant scanning rate of 10 K/min.

Dielectric measurements at frequencies from 0.1 Hz up to 10 MHz were carried out using a frequency-response analyzer (Novocontrol Alpha). For $X$ = MN and PN, additional high-frequency measurements up to about 2 GHz were performed employing an impedance analyzer (Keysight E4991B) with coaxial reflectometric setup.[35] The sample materials were heated to ensure a rather low-viscous liquid state and filled into preheated parallel-plate capacitors with plate distances of 0.1 mm. For cooling and heating, a $N_2$-gas cryostat was used. Within the cryostat, before each measurement run the samples were first sufficiently heated to ensure a fully liquid state. The measurements were done under cooling.

## III. RESULTS AND DISCUSSION

### A. DSC results

Figure 1 shows the DSC results for the three investigated mixtures as measured with a cooling/heating rate of 10 K/min. For $X$ = MN [Fig. 1(a)], the sharp negative (i.e., exothermic) peak, observed under cooling at about 305 K, indicates the transition from the liquid into the PC phase. The latter can be easily supercooled and at low temperatures a so-called glassy crystal state is formed where the reorientational motions exhibit glassy freezing. This is signified by the anomalies observed somewhat around 150 K under cooling and heating, which reveal the typical signature of a glass transition and are especially well pronounced for the heating curve [inset of Fig. 1(a)]. The orientational glass transition occurs at $T_g^o \approx 149$ K (using the onset evaluation method of the heating curve). Such glass-like freezing of the reorientational degrees of freedom of the molecules is a well-established phenomenon and found in many PCs.[36,37,38,39,40] The melting peak observed during heating at about 310 K is more smeared out than under cooling and located at higher temperature as was also found for the SN-GN system (see supplementary information of Ref. 16). Overall, this DSC trace for $X$ = MN closely resembles that of the corresponding $SN_{0.85}GN_{0.15}$ mixture[16] and reveals a broad stability range of the PC phase, without any interference from a transition to a more ordered crystalline phase as found in pure SN.[25]

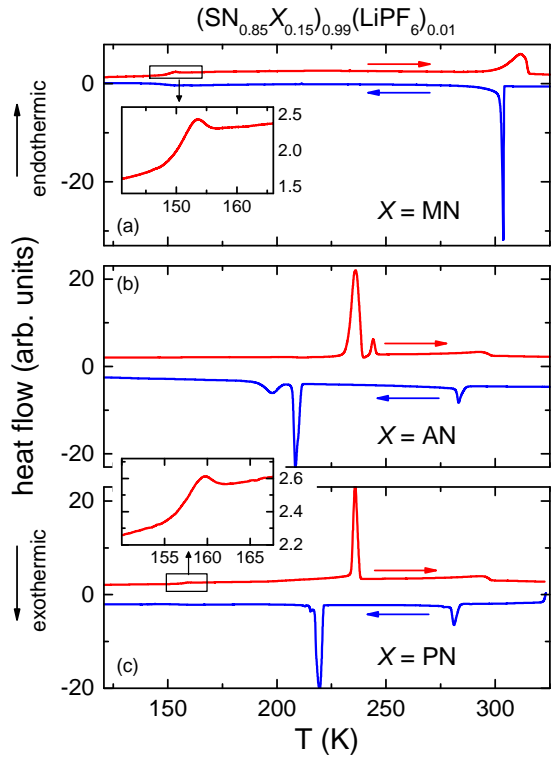

**FIG. 1.** DSC heat flow as measured under cooling (lower lines) and heating (upper lines) with rates of 10 K/min for the three investigated mixtures. Endothermic processes are plotted in positive $y$-direction. The insets show zoomed views of the glass transition during heating observed for $X$ = MN and PN.

In contrast to SN-MN, for the SN-AN mixture [Fig. 1(b)], an additional exothermic peak at cooling (at 208 K) and the corresponding endothermic one at heating (236 K) show up. As no indication for a glass-transition anomaly appears at lower temperatures, this peak marks a transition from the PC to the orientationally ordered crystal. Nevertheless, the PC phase can at least be studied at temperatures above this transition. The main DSC peak caused by the orientational order is accompanied by a second, smaller peak. Currently it is not clear whether it may indicate a second transition (e.g., from a phase with restricted orientational motion to full orientational order) or partial phase separation at low temperatures in this mixture.[23] Notably, the amplitudes of the crystallization/melting anomalies, observed at about 283 and 294 K for SN-AN [Fig. 1(b)], seem to be significantly smaller that for SN-MN [Fig. 1(a)]. Obviously, in this mixture the main entropy reduction under cooling arises at the formation of orientational ordering and not at the liquid-crystal transition. Recently it was pointed out that such large entropy changes, as often found at the orientational-order transition in PCs, may be used to develop new, environmentally friendly refrigeration techniques.[41,42]

For $X$ = PN [Fig. 1(c)], the DSC trace in principle resembles that of the SN-AN mixture with crystallization and melting peaks at about 281 and 296 K, respectively, and an additional, much stronger ordering transition at 220 and 236 K for cooling and heating, respectively. However, an inspection of the low-temperature region reveals a significant glass-transition anomaly at about 156 K [inset of Fig. 1(c)]. This indicates that in this mixture the second phase transition found under cooling only involves partial orientational ordering and restricted reorientational motions (e.g., about a single axis only) still are possible, finally undergoing glassy freezing. Such transitions, where the reorientational motion changes from almost isotropic to anisotropic, are well documented for various other PCs.[19,43] An alternative explanation of the observed glass transition would be a phase separation into PC regions and regions of orientational order[23] (this issue will be discussed in more detail below). In any case, it is interesting to check how much the ionic charge transport is affected by the transition into this partially ordered state.

### B. Dielectric spectra

#### 1. SN-MN

Figure 2 shows spectra of the dielectric constant $\varepsilon'$ (a), dielectric loss $\varepsilon''$ (b), and conductivity $\sigma'$ (c) of the SN-MN system for selected temperatures. The measurements were performed during cooling from room temperature with 0.4 K/min. With increasing frequency, the $\varepsilon'$ spectra [Fig. 2(a)] reveal a step-like decrease from about 30 to 2, which shifts to lower frequencies with decreasing temperature. It is accompanied by a peak in the loss spectra [Fig. 2(b)]. These are the typical signatures of a relaxational process.[44,45,46] It can be ascribed to the so-called $\alpha$ relaxation, i.e., the global reorientational motions of the PC molecules. The mentioned shift to lower frequencies mirrors the continuous slowing down of this molecular dynamics under cooling, typical for the glass transition[44,45] and found in various other PCs.[19] From the loss-peak frequencies $\nu_p$, the average relaxation time $\langle \tau \rangle$ of the $\alpha$ process can be estimated via $\langle \tau \rangle \approx 1/(2\pi\nu_p)$. Notably, for 157 K, the lowest temperature where the loss peak is fully seen in the frequency window of Fig. 2(b), the amplitudes of the relaxation features in $\varepsilon'$ and $\varepsilon''$ are significantly reduced and the loss peak exhibits a more symmetric shape than at the higher temperatures. This may well reflect an at least partial transition into a more ordered state, not detected in the DSC experiments [Fig. 1(a)] because they were performed with a higher cooling rate of 10 K/min.

For the three highest temperatures included in Fig. 2(a), $\varepsilon'(\nu)$ exhibits a strong additional increase for frequencies below those of the $\alpha$-relaxation step and the dielectric constant approaches unreasonably high ("colossal"[47]) values beyond 1000. This can be ascribed to electrode polarization, a phenomenon commonly found for ionic conductors (also termed "blocking electrodes"). It arises at low frequencies and relatively high temperatures when the mobile ions arrive at the metallic electrodes which leads to the formation of thin, poorly conducting space-charge regions that act as a huge capacitor.[48] At sufficiently high frequencies, the capacitor becomes effectively shorted and the intrinsic response is detected.[47,48]

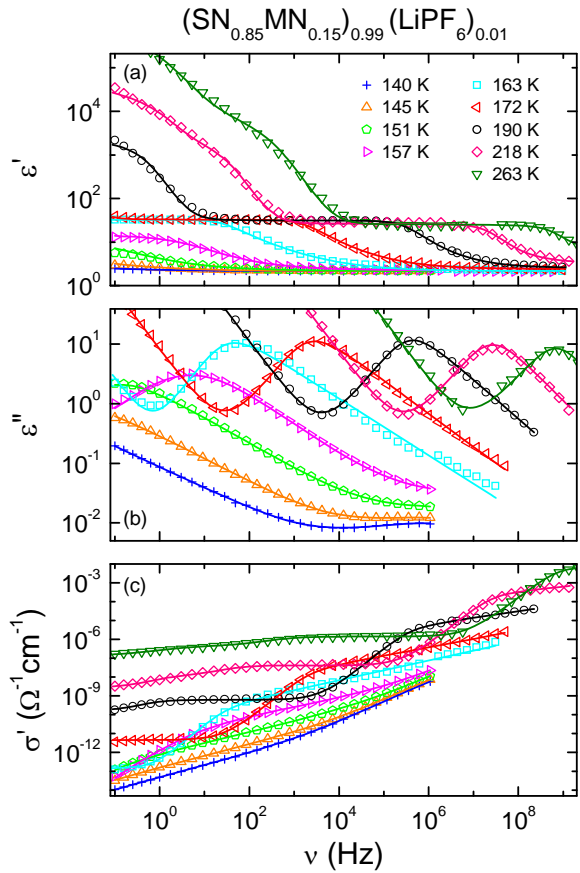

**FIG. 2.** Frequency-dependent dielectric constant (a), dielectric loss (b), and conductivity (c) of the SN-MN mixture for various temperatures. The solid lines in (a) and (b) are fits of $\varepsilon'$ and $\varepsilon''$ using an equivalent-circuit approach[48] consisting of up to two distributed RC circuits to describe the electrode-polarization effects, a dc-conductivity contribution, and up to two relaxation functions (see text for details). The lines in (c) were calculated from the fits of $\varepsilon''$.

In the loss spectra [Fig. 2(b)], at frequencies below the peaks, a marked increase is found. Its slope of -1 in the double-logarithmic plot implies $\varepsilon'' \propto \nu^1$ in this region. Via the relation $\sigma' \propto \varepsilon''\nu$, this corresponds to constant conductivity. Consequently, in Fig. 2(c) a frequency-independent plateau precedes the shoulders at high frequencies that correspond to the loss peaks in Fig. 2(b). For the three highest temperatures, at low frequencies $\sigma'(\nu)$ deviates from this plateau and becomes reduced, which is due to the blocking-electrode effects mentioned above.[48]

In addition to the right flank of the $\alpha$-relaxation peak, in Fig. 2(b) the loss spectra at the lowest temperatures reveal the indication of a second, broad peak at high frequencies. The occurrence of such secondary relaxation peaks is typical for glass-forming liquids[49,50,51] and also found for certain PCs,[19] including various SN-GN mixtures.[16,25,26] A detailed treatment of this feature is out of the scope of the present work.

To describe the spectra of Fig. 2, we applied an equivalent-circuit approach, simultaneously fitting $\varepsilon'$ and $\varepsilon''$.[47,48] For $\sigma'$, the fit curves were calculated via $\varepsilon'' = \sigma'/(2\pi\nu\varepsilon_0)$ where $\varepsilon_0$ is the permittivity of free space. As shown by us in various previous works,[16,31,32,47,48] electrode-polarization can be formally well modeled by distributed RC circuits, assumed to be connected in series to the bulk.[48] The electrode-dominated regions in Figs. 2(a) and (c) seem to exhibit two successive steps. Such behavior is often found in ionic conductors and possible microscopic origins were discussed, e.g., in Ref. 48. Therefore, for the highest temperatures two RC circuits were employed to account for the electrode effects. For the $\alpha$ relaxation we used the empirical Cole-Davidson function[52], well known to provide a reasonable description for glass formers and PCs.[19,45,53] The secondary relaxation was described by the Cole-Cole function[54] which is common for such processes.[19,51] The contribution of the dc conductivity to the loss is given by $\varepsilon''_{dc} = \sigma_{dc}/(2\pi\nu\varepsilon_0)$. It should be noted that, depending on temperature, only part of these elements were needed (e.g., at low temperatures, the electrode effects are outside of the frequency window), which kept the number of parameters at a reasonable level. The fits obtained in this way (lines in Fig. 2) reasonably agree with the experimental data. The dc-conductivity and relaxation times deduced by the fits perfectly match the values read off from the spectra.

### 2. SN-PN

The dielectric spectra for the SN-PN mixture, presented in Fig. 3, in principle show the same contributions as those for SN-MN. The main difference is the discontinuity found between 230 and 218 K: for $T \leq 218$ K, the relaxation amplitude (step height in $\varepsilon'$ and peak area in $\varepsilon''$) is strongly reduced by nearly one decade. In light of the above discussion of the DSC results [Fig. 1(c)], this most likely reflects the transition from the "classical" PC state with isotropic reorientations to restricted reorientational motions. Qualitatively similar behavior was also reported for other PCs, e.g., ortho-carborane[43] and adamantanone.[19] The assumption that this transition does not lead to complete orientational order, concluded above from the occurrence of a glass transition in the DSC trace, is corroborated by the finding of a clear relaxation process even below the transition. The data of Fig. 3 were fitted in the same way as those of Fig. 2 (lines) which, again, works reasonably well. Some minor deviations of experimental data and fits show up in the crossover region between electrode polarization and $\alpha$ relaxation in Fig. 3(a). This may be due to another relaxation process (intrinsic or non-intrinsic) or ac conductivity as discussed in some detail in Refs. 23 and 25 for SN-GN mixtures. To avoid a too excessive number of parameters, we refrain here from adding another element to the employed fit function. This has no significant influence on the results for the $\alpha$-relaxation time and the dc conductivity, which are the only relevant parameters for the further discussion. The dc conductivity, estimated by the frequency-independent plateau in $\sigma'(\nu)$ [Fig. 3(c)] does not seem to be strongly affected by the phase transition, a fact that will be discussed in more detail below.

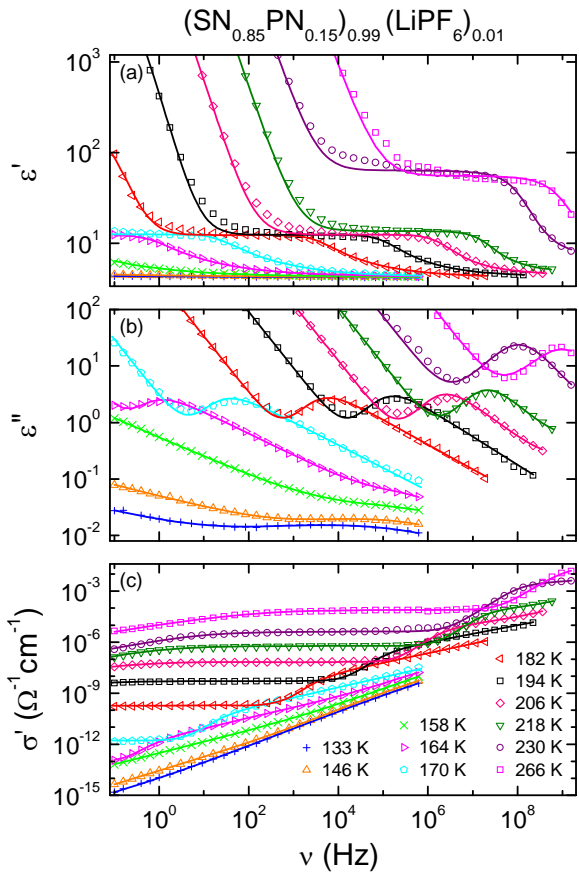

**FIG. 3.** Frequency-dependent dielectric constant (a), dielectric loss (b), and conductivity (c) of the SN-PN mixture for various temperatures. The lines have the same meaning as in Fig. 2.

### 3. SN-AN

For the SN-AN mixture, no relaxational response is observed within the investigated frequency window (Fig. 4). At temperatures $T \geq 221$ K, i.e. above the transition to complete orientational order, the intrinsic dielectric-constant plateau detected at high frequencies is rather high, pointing to the presence of relaxation steps at higher frequencies as also found in the other systems (Figs. 2 and 3). If this relaxation should persist below the transition, it should shift into the frequency window under cooling as observed for the PN sample (Fig. 3). However, at 194 and 206 K in Fig. 4(a), $\varepsilon'(\nu)$ is rather featureless and low. This well corresponds to the complete order below this transition, suggested above based on the DSC results. This finding corroborates the notion that the reorientations in PCs are essential for their high ionic conductivity. The spectra of Fig. 4 were fitted just as for the other two systems but without the $\alpha$ relaxation (lines). At the two lowest temperatures, the conductivity is low and no element for electrode polarization was necessary. There, for the intrinsic response an additional power law, $\sigma' = \sigma_{dc} + \sigma_0 \nu^s$ (and $\sigma'' = \tan(s\pi/2)\nu^s$ according to the Kramers-Kronig relation[55]) was used, termed universal dielectric response (UDR) and commonly ascribed to hopping conductivity.[55,56,57,58,59] It leads to a better description of the small residual conductivity in this ordered phase than pure dc conductivity alone. Significant information on the dc conductivity could only be obtained for 206 K.

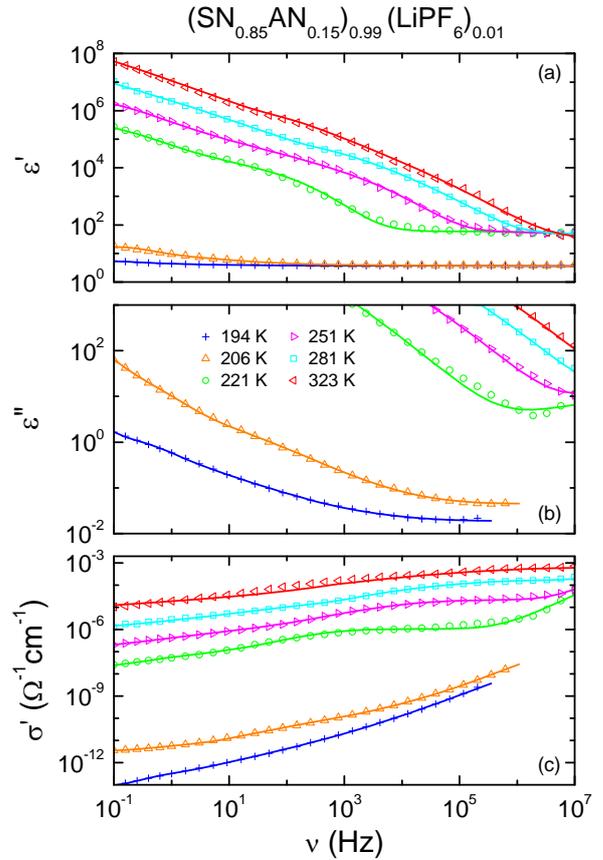

**FIG. 4.** Frequency-dependent dielectric constant (a), dielectric loss (b), and conductivity (c) of the SN-AN mixture for various temperatures. The lines are fits as in Figs. 2 and 3 but without the $\alpha$ relaxation. For the two lowest temperatures, an additional UDR contribution was assumed.

### C. Dc conductivity

Figure 5 shows the dc conductivity of all investigated mixtures, plotted versus the inverse temperature in an Arrhenius representation. The melting temperatures and the temperatures of the transition into a more ordered crystalline phase, deduced from the DSC experiments under cooling (Fig. 1), are indicated by vertical lines. Just as previously reported for the corresponding SN-GN mixture (shown by the squares in Fig. 5),[16] the dc conductivity for all mixtures investigated in the present study is significantly higher than for the pure SN doped with the same percentage of Li salt (inverted triangles; this data set was remeasured here up to higher temperatures than previously reported in Ref. 16). At high temperatures in the PC phase, $\sigma_{dc}$ is found to considerably increase with increasing molecule size. For the two largest molecules (PN and AN) this increase becomes smaller and the highest conductivity is detected for the $X$ = PN mixture. For example at 280 K, it exceeds that of $SN_{0.99}(LiPF_6)_{0.01}$ by about 2.5 decades, again demonstrating the very strong influence of admixing on the ionic conductivity.[16,22] Interestingly, while SN without admixture reveals a very strong difference of $\sigma_{dc}$ in the liquid and PC phase, the jump of $\sigma_{dc}(T)$ at the transition becomes successively smaller for larger admixture-molecules and nearly vanishes for PN and AN. One key for reaching the highest conductivities in these systems obviously is the conservation of a liquidlike conductivity at the crossover into the PC phase. This is confirmed by the similar behavior found in our previous investigation of the SN-GN system with different GN concentrations:[16] The highest $\sigma_{dc}$ values were detected for the highest GN concentrations, which had the least pronounced conductivity anomalies at crystallization.

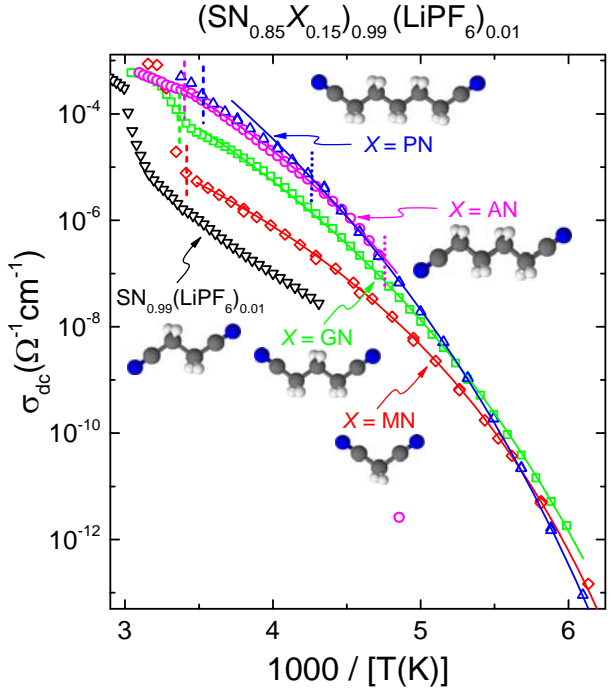

**FIG. 5.** Temperature-dependent dc conductivity of the investigated $SN_{0.85}X_{0.15}$ mixtures with 1% $LiPF_6$ shown in an Arrhenius representation. In addition, results for the corresponding SN-GN mixture from Ref. 16 and for SN without another nitrile are shown. The molecules are schematically indicated. The solid lines are fits with the VFT equation for the conductivity, Eq. (1). The vertical dashed lines indicate the melting points. For PN and AN, the vertical dotted lines show the transition temperatures into a more ordered phase (cf. DSC measurements in Fig. 1). Note the single data point below the orientational-ordering transition in SN-AN at about $3\times10^{-12}$ $\Omega^{-1}cm^{-1}$.

As discussed above, for $X$ = PN and AN, aside of the crystallization, a second transition shows up at lower temperatures (cf. Fig. 1), indicated in Fig. 5 by the vertical dotted lines. While for the SN-PN mixture only a small change of slope $\sigma_{dc}(T)$ is found at this transition, for $X$ = AN the dc conductivity becomes strongly reduced by many decades [only a single data point immediately below the transition could be obtained; cf. Fig. 4(c)]. This finding is well consistent with the above conclusions from the DSC results: At this transition, the SN-PN mixture transfers into a phase with restricted reorientational motions while the SN-AN systems undergoes full orientational order. The still possible reorientations for $X$ = PN, albeit restricted, obviously still provide sufficient freedom for the ions to pass the molecules. In contrast, the fully ordered state for $X$ = AN, lacking any molecule rotations, no longer supports any sizable ionic charge transport.

The almost unaffected conductivity below the low-temperature transition in SN-PN (upright triangles in Fig. 5) allows to essentially exclude the alternative explanation of the detected glass transition in terms of phase separation, which was considered in the above discussion of Fig. 1(c): If the strong anomaly in the DSC trace would mark a transition into a predominantly orientationally ordered, highly insulating phase with small, phase-separated PC regions, it seems unlikely that this transition should have such little effect on the conductivity.

The non-linear behavior of $\log(\sigma_{dc})$ vs. $1/T$ observed in the PC phases of all mixtures (Fig. 5) evidences clear deviations from Arrhenius behavior, $\sigma_{dc} \propto \exp[-E/(k_BT)]$, expected from plain thermal activation over an energy barrier $E$. Such non-Arrhenius temperature dependence of the conductivity is a characteristic property of ionically conducting glasslike materials, at least if some coupling of the ionic mobility to the glassy dynamics prevails.[8,16,24,31,32,60,61] In structural glass formers, the latter mainly involves the translational motions of the molecules, atoms, etc. (having in mind the original definition of the glass transition in terms of viscosity), while in PCs the glassy dynamics is given by the reorientational motions. Non-Arrhenius behavior of dynamic quantities in glass-forming materials usually can be well described by the empirical Vogel-Fulcher-Tammann (VFT) law.[62,63,64,65] For $\sigma_{dc}(T)$ in Fig. 5 we use its variant for the conductivity:

$$\sigma_{dc} = \sigma_0 \exp\left[\frac{-DT_{VF}}{T - T_{VF}}\right] \qquad (1)$$

Here $\sigma_0$ is a pre-exponential factor, $D$ is the so-called strength parameter quantifying the deviation from Arrhenius behavior[65] and $T_{VF}$ is the Vogel-Fulcher temperature where $\sigma_{dc}$ would diverge. As shown by the solid lines, $\sigma_{dc}(T)$ in Fig. 5 can be well fitted in this way. For $X$ = PN, the temperature region of the PC phase with isotropic reorientations (between the corresponding dashed and dotted vertical lines in Fig, 5) is too small to allow for meaningful statements about non-Arrhenius behavior of $\sigma_{dc}$. However, in the low-temperature phase a fit with a VFT law (line through the upright triangles in Fig. 5) works reasonably well, indicating that here glasslike dynamics still exists and dominates the charge transport. This is consistent with the finding of a glass transition in the DSC trace found in this phase [Fig. 1(c)].

### D. Comparing ionic conductivity and orientational dynamics

In Ref. 16 the strong increase of the dc conductivity of the SN-GN mixture with increasing GN content was rationalized by a successively better coupling of the reorientational and translational dynamics for higher concentrations of the larger GN molecule. For the present SN-MN and SN-PN samples, this coupling is examined in Figs. 6(a) and (b). They provide a comparison of the temperature dependences of the average reorientational relaxation times $\langle\tau\rangle$ and the inverse dc conductivity (the dc resistivity $\rho_{dc} = 1/\sigma_{dc}$) by plotting both quantities in a common frame using an Arrhenius representation. [Both, $\langle\tau\rangle$ determined by reading off the loss-peak frequencies and from the fits performed for selected spectra (lines in Figs. 2 and 3) are shown. The $\sigma_{dc}$ values also were either read off from the $\sigma'$ spectra or determined by the fits.] By a proper choice of the ordinate starting values, and ensuring the same number of decades per scale unit for the $\rho_{dc}$ and $\langle\tau\rangle$ ordinates, it is possible to achieve a good match of both data sets. This is the case for both $X$ = MN and PN, in marked contrast to $X$ = GN shown in Fig. 6(c).[16]

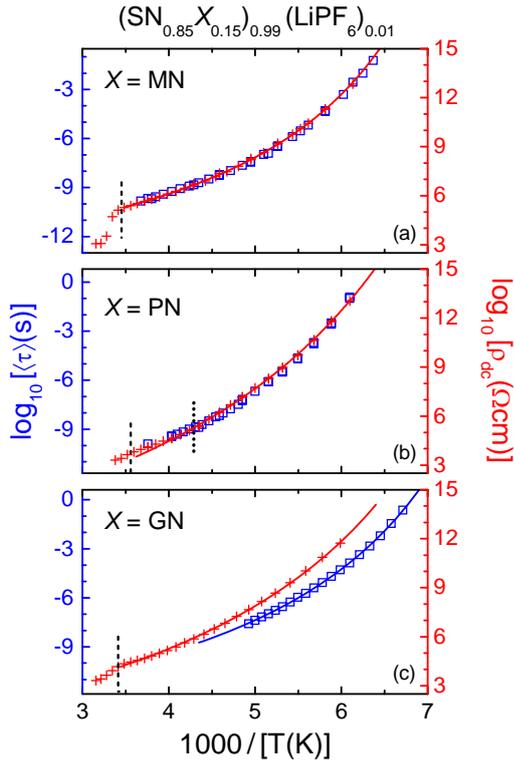

**FIG. 6.** Temperature dependences of the average reorientational relaxation time and of the dc resistivity of the SN-MN (a), SN-PN (b), and SN-GN (c) mixtures (the latter from Ref. 16) with 1 mol% LiPF$_6$, plotted in Arrhenius representation. The axis ranges of $\langle\tau\rangle$ (squares; left axes) and $\rho_{dc}$ (plusses; right axes) were adjusted to cover the same number of decades. The starting values of the ordinates were chosen to achieve a match of both quantities at high temperatures in the PC phases (extrapolated for SN-GN). The vertical dashed lines indicate the liquid-PC transitions. For $X$ = PN the vertical dotted line denotes the transition into the suggested phase with higher orientational order. The solid lines through the $\rho_{dc}(T)$ data are VFT curves calculated from the fits of $\sigma_{dc}(T)$ shown in Fig. 5. The $\langle\tau\rangle(T)$ data in frame (c) were fitted by a separate VFT law.[65]

The good scaling of $\langle\tau\rangle$ and $\rho_{dc}$, revealed in Fig. 6 for SN-MN and SN-PN, indicates proportionality of both quantities evidencing a close coupling of the ionic translational to the molecular reorientational dynamics. This strongly points to the relevance of the revolving-door mechanism for the ionic mobility in these mixtures. In Ref. 16, the conductivity enhancement when adding the larger GN to the smaller SN molecules was proposed to be due to a more effective revolving-door mechanism, mirrored by increasingly better rotational-translation coupling with increasing GN content. Comparing $X$ = GN and PN in Fig. 6, it thus seems reasonable that admixing an even larger molecule (PN) to SN may lead to even better coupling and to a further enhancement of the conductivity (Fig. 5). It should be noted that nearly perfect coupling as found for SN-PN at 15 mol% PN content, for SN-GN is only reached at much higher concentrations of 80 mol%.[16] For this high GN concentration, the high-temperature conductivity in the PC phase is of similar magnitude as for the present 15 mol% SN-PN mixture.

In light of the above considerations, at first glance it does not seem plausible that such nearly perfect coupling of $\langle\tau\rangle$ and $\rho_{dc}$ is also revealed by the SN-MN system where 15 mol% of a *smaller* molecule is admixed [Fig. 6(a)]. However, this system may of course also be regarded as a 85% admixture of the larger GN molecule to MN and the situation may be similar as for the SN$_{0.2}$GN$_{0.8}$ mixture investigated in Ref. 16 where nearly perfect coupling was found, too. It seems that large amounts of a somewhat larger molecule (SN$_{0.2}$GN$_{0.8}$ and SN$_{0.85}$MN$_{0.15}$) or small amounts of a much larger molecule (SN$_{0.85}$PN$_{0.15}$) both can lead to very strong coupling of ion mobility and molecular rotation. However, despite its equally good coupling, the SN-MN mixture exhibits much less enhanced conductivity than SN-PN (Fig. 5). This can be explained by the smaller average molecule size in the SN-MN mixture, which should lead to closer packing, even when the molecules are rotating. This results in less overall free volume available for the ions, partly compensating the more effective revolving-door mechanism in the mixture compared to unmixed SN and, thus, explains the found smaller conductivity enhancement.

The close coupling of relaxation time and dc resistivity reminds of the empirical Barton-Nakajima-Namikawa (BNN) relation, originally reported for ionically conducting glasses.[66,67,68] It predicts proportionality of the dc conductivity and the rate of a relaxation process that arises from ionic diffusion in these glasses. In contrast, the relaxation process detected in the present PC samples obviously is of reorientational origin.

Finally, it should be noted that the markedly different dc conductivities of the investigated SN-$X$ mixtures, observed in the PC phase at high temperatures (Fig. 5), is not caused by strongly different reorientational relaxation times. This is revealed by Fig. 7, showing the temperature dependences of the reorientational $\alpha$-relaxation times for $X$ = MN, GN,[16] PN, and SN without admixed nitrile.[16] In the temperature range where $\sigma_{dc}$ varies by up to 2.5 decades depending on the admixed molecule (above about 200 K; Fig. 5), $\langle\tau\rangle$ of the different materials is of comparable magnitude and cannot explain the findings of Fig. 5. Instead, the translation-rotation coupling, concentration, and size of the involved molecules determines the absolute values of $\sigma_{dc}$ for a given rotation rate of the revolving doors.

As revealed by Fig. 7, for lower temperatures the relaxation times of the different mixtures no longer are similar but differ significantly. This obviously arises from the different orientational glass temperatures of these systems. They can be estimated from extrapolating the VFT fits shown as lines in Fig. 7 to a value of $\langle \tau \rangle = 100$ s, usually found at the glass transition. We obtain 149 and 155 K for MN and PN, respectively. This well matches the anomalies found in their DSC traces (insets of Fig. 1). As $\langle \tau \rangle$ of the investigated materials strongly differs at low temperatures, in this region, just as in the previously investigated mixtures of cyclohexanol and cyclooctanol,[22] strong effects of $\langle \tau \rangle$ on the conductivity can be expected. Indeed with decreasing temperature, the $\sigma_{dc}$ curves of the materials in Fig. 5 approach each other and finally cross. At the lowest temperatures (e.g., at $1000/T = 6$, corresponding to about 167 K) the influence of the strongly different $\langle \tau \rangle$ values obviously dominates over the other effects and leads to the highest conductivity for the GN system and the lowest for PN (Fig. 5). This is in accord with the succession of relaxation times revealed at low temperatures in Fig. 7, the fastest reorientations leading to the highest conductivity.

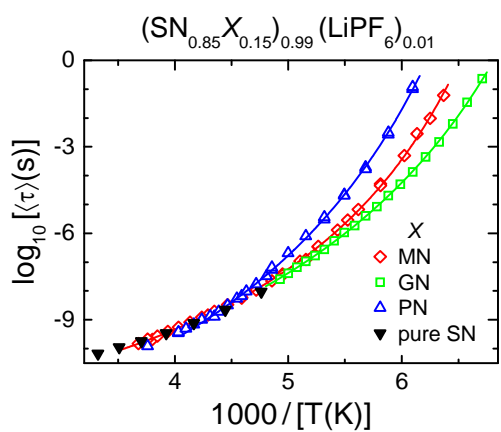

**FIG. 7.** Comparison of the average $\alpha$-relaxation times of the mixtures shown in Fig. 6 and of SN without admixed nitrile. The data for the latter and for SN-GN were taken from Ref. 16. The lines are fits with the VFT law.[65]

## IV. SUMMARY AND CONCLUSIONS

In summary, using dielectric spectroscopy we have investigated the translational ionic and reorientational molecular dynamics of three mixtures of plastic-crystalline SN with related nitriles of different molecule size. Just as previously reported for SN-GN, all these mixtures show a significant enhancement of their ionic conductivity if compared to the pure SN system which can reach up to 2.5 decades. One of these mixtures (SN-MN) exhibits a PC phase that extends to the lowest investigated temperatures and reveals glassy freezing of the orientational degrees of freedom. In SN-PN, a transition to restricted reorientational motions is observed, which, however, only has a minor effect on the high conductivity of this system. Only SN-AN undergoes a transition into complete (translational and orientational) crystalline order under cooling below 208 K, where the conductivity is strongly suppressed.

Both the addition of 15 mol% of smaller (MN) or larger molecules (PN and AN) to SN induces an increase of the conductivity at high temperature, close to the liquid state. A comparison of ionic dc conductivity and reorientational molecular relaxation times, possible for MN and PN, reveals good coupling of both dynamics for both cases. This further supports the notion that the rotating molecules advance ionic mobility in a revolving-door like mechanism. Nevertheless, at high temperatures different absolute values of the conductivity are found for SN-MN and the two other systems despite they feature similar relaxation times and similar coupling. Obviously, even for good translation-rotation coupling, the same reorientational dynamics of the molecules can lead to very different dc conductivities. A key for reaching high conductivities in the solid state is the avoidance of a significant conductivity jump at the transition from the liquid to the PC phase, thus preserving liquidlike conductivity below the melting point. In the present case, this is reached by admixing rather small amounts of much larger molecules than those of the host system. Further work, investigating additional concentrations of admixed nitriles and salts, and also checking the behavior for different nitriles and other salts than $LiPF_6$,[2] should be performed to examine all options of optimizing SN as a plastic-crystalline solid-state electrolyte. Here an important task certainly is also the shift of the liquid-PC phase transition to higher temperatures.

Together with our previous results on different SN-GN and cyclohexanol-cyclooctanol mixtures,[16,22] the present investigation reveals the parameters that are relevant for the conductivity enhancement in PC mixtures: the size of the added molecules, the degree of translation-rotational coupling, the concentration ratio of large to small molecules, and the molecular reorientation rate at a given temperature. Optimizing these four factors is essential for the development of mixed PC electrolytes for future electrochemical applications.

## ACKNOWLEDGEMENTS

This work was supported by the Deutsche Forschungsgemeinschaft (grant No. LU 656/3-1).